\documentclass[12pt,letterpaper]{JHEP3}
\pdfoutput=1
\usepackage{epsfig}
\usepackage{subfigure}
\usepackage{cite}

\graphicspath{{Figures/}}

\title{Complementarity + back-reaction is enough}

\author{Lam Hui\footnote{lhui@astro.columbia.edu} \ 
and I-Sheng Yang\footnote{isheng.yang@gmail.com}\\
$^*$ ISCAP and Physics Department \\
Columbia University, New York, NY, 10027 , U.S.A. \\
$^\dagger$ IOP and GRAPPA, Universiteit van Amsterdam, \\
Science Park 904, 1090 GL Amsterdam, Netherlands
}

\abstract{We investigate a recent development of the black hole information problem, in which a practical paradox has been formulated to show that complementarity is insufficient. A crucial ingredient in this practical paradox is to distill information from the early Hawking radiation within the past lightcone of the black hole. By causality this action can back-react on the black hole. 
Taking this back-reaction into account, the paradox could be resolved without invoking any new physics beyond complementarity. 
This resolution requires a certain constraint on the $S$-matrix to be satisfied.
Further insights into the $S$-matrix could potentially be obtained
by effective-field-theory computations of the back-reaction on the nice slice.
}

\begin{document}

\section{Introduction}
\label{intro}

The black hole information paradox \cite{Haw76a} has always been an inspiring topic. Recent arguments made by Almheiri, Marolf, Polchinski and Sully (AMPS) \cite{AMPS} 
(see also \cite{BraPir09}) led to a new surge of discussions\footnote{See \cite{AMPSS} for an extensive list of references.}.
They argued that complementarity \cite{Tho90b,SusTho93,Sus93,Ver95,LowPol95,Bou09}, a conjecture previously accepted by most, was not enough to resolve the information paradox. In this paper, we wish to explore the possibility that complementarity is actually
sufficient to resolve the problem, once back-reaction
(from the maniplulation of Hawking quanta necessary in practical versions of the paradox)
is taken into account. 
%Most responses to AMPS pick a side. Either one accepts the paradox and envision some new physics, or one tries to find loopholes in the argument. Both are worthy endeavors. In this paper we will take the second stance. The claim is that the firewall paradox can be resolved within the framework of complementarity. We will demonstrate how, and also point out possible new physics we can learn from such a conservative resolution.

\begin{figure}[ht]
\begin{center}
\includegraphics[width=12cm]{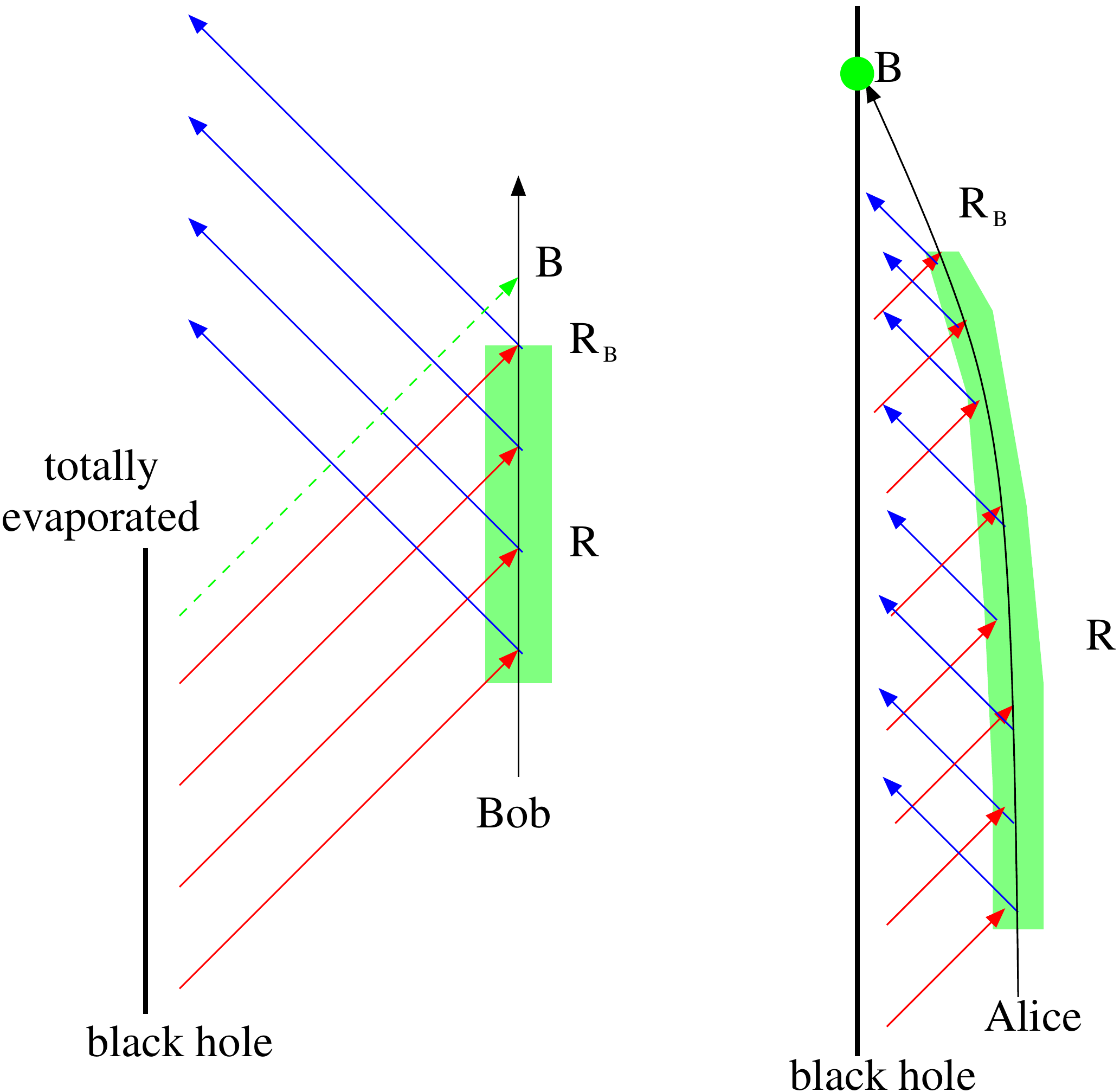}
\caption{Black holes are drawn as thick black lines, with their near-horizon and interior details suppressed (which includes the interior mode $A$). The left figure shows Bob, who stays very far away. His distiller (operating in the green box) is space-like separated from the near-horizon origin of the late quantum $B$, which then propagates to Bob along the green dotted line. Back-reaction from his distiller is irrelevant but he also cannot jump into the black hole. The right figure shows Alice, who stays near the black hole since she wants to jump in later. The back-reaction from her distiller will affect the late quantum $B$ (the green dot) she encounters later.
\label{fig-intro}}
\end{center}
\end{figure}

Le us first review the information problem.
Throughout this paper, we adopt the notation where
$M_{\rm Planck}=1$, so $M$ stands for both the mass and also 
the horizon size of the black hole (or more precisely, half
the Schwarzschild radius).
Let $B$ be a particular near-horizon mode after Page time \cite{Pag93}, $A$ be its interior partner, $R$ be the early Hawking radiation. Knowing the initial state, the unitary evaporation allows us to distill a minimal subsystem $R_B$ from $R$ such that it is maximally entangled with $B$. On the other hand, the equivalence principle demands that $A$ and $B$ are maximally entangled as the in-falling vacuum state. These two facts put together violate the monogamy of entanglement. A possible solution is $A=R_B$, or colloquially ``in = out'', which means that the interior is identified with the early Hawking radiation. 

This follows the spirit of complementarity in that the interior and exterior cannot be treated as two independent sets of degrees of freedom. Note that it should only work if $A$ and $R_B$ can never be brought together and compared even in principle, for example if Bob distills $R_B$ from $R$ at distance $\sim M^3$ away as shown in Fig.\ref{fig-intro}-left. Here, ``in = out'' can be thought of as a consolation to those who insist on thinking globally. As a practical matter, since ``in'' and ``out'' can never be brought together and compared, by definition it does not lead to any paradox.

However the key argument in \cite{AMPS} is to point out that there is a practical paradox as shown in Fig.\ref{fig-intro}-right.
%\footnote{Despite the common habit to call \cite{AMPS} the firewall paper, the key point is showing that complementarity is not enough. The firewall phenomenon was already pointed out earlier under the name of energy curtain \cite{BraPir09} and is not directly relevant to the discussion here.}. 
After leaving the black hole, Hawking radiation is basically free-streaming. It should not make any difference if someone intercepts them earlier. 
Alice intercepts the radiation earlier, and closer to the black hole. She thus has
enough time to distill $R_B$ from $R$, and then carry it into the black hole to compare with $A$. $A=R_B$ now becomes a blatant quantum cloning. From our point of view, this practical paradox is the strongest version of the AMPS argument. One single observer, Alice, conducts two experiments and sees conflicting results. 
Any proposed resolution must directly confront this practical paradox.
\footnote{Note that for a practical paradox, it is essential to distill $R_B$, or at least to have a system smaller than $R$ that entangles with $R_B$ (such as performing a classical measurement). That is because Alice cannot bring all $\sim M^2$ qubits in $R$ back into the black hole without dramatically altering the geometry. She must process the information in $R$ one way or another to reduce the amount of information she needs to carry.}

First let us revisit Fig.\ref{fig-intro}. Comparing Bob and Alice,
Bob performs distillation further away and later, while Alice
does the same closer in and earlier. The fact that the distillation
is done at different times does not greatly affect the state
of $R$ received by them -- $R$ earlier and later are simply
related by the unitary transformation associated with free streaming.
There is, however, one crucial difference in the distillation
processes performed by Bob and Alice, due to their space-time locations
in relation to the black hole.
By causality, Bob's action cannot affect the black hole in any way, and there is no paradox for him. On the other hand, Alice can jump into the black hole to witness a potential paradox, but {\it the same fact ensures that her actions can causally affect the black hole}. If complementarity can survive without 
additional new physics, the key must be in how the distillation process done by Alice back-reacts on the black hole. 
%Our goal in this paper is to explore the possibility of such a causal back-reaction,
%and how this might resolve the paradox.

The resolution we wish to explore can thus be described as ``complementarity + back-reaction'':
1) complementarity, in the sense of ``in=out'', addresses situations in which the distillation process
is space-like separated from the near-horizon origin of quantum $B$; 2) back-reaction on the black hole addresses situations where the near-horizon origin of quantum $B$ is within the forward
light-cone of the distillation process. Our main goal in this paper is to demonstrate why
such a back-reaction is plausible, and how it could resolve the paradox.

The idea of back-reaction is in a sense a natural one, but there has not been
much discussion in the literature on how this could address the AMPS paradox.
There are probably several reasons for this.

First, a perhaps common response to the idea of back-reaction is to say, yes, in principle
this could happen, but presumably a clever experimentalist can make the
back-reaction sufficiently weak to be negligible. We will argue that
the fairly non-trivial distillation process, whereby Alice obtains the complement of $B$ from $R$,
necessarily gives rise to a certain level of back-reaction. Tuning the back-reaction to
be acceptably small could sacrifice one's ability to distill $R_B$ from $R$.

Second, one might think that since back-reaction involves sending signals
from outside into the black hole, it seems to go in the wrong direction i.e.
opposite direction from the late outgoing Hawking quantum $B$ which we wish
to affect. This isn't an issue, as long as one keeps in mind that the back-reaction
could affect the state of the black hole and therefore its emission as well.

Third, one might think some kind of shield can be set up to
prevent back-reaction signals from reaching the black hole.
From Fig.\ref{fig-intro}-right, it is apparent that setting up a shield
to block off a significant fraction of the back-reaction signals would
also prevent a fraction of the early Hawking radiation from reaching
the distillation apparatus. 

%This idea is not new, but previous discussions seemed to be stopped short by two common perceptions against back-reaction --
%\begin{itemize}
%\item It is tunable: \\
%By designing a better apparatus, the back-reaction can be arbitrarily reduced.
%\item It goes in the wrong direction: \\
%Back-reaction involves sending signals from outside into the black hole, which should have crossed the horizon and crashed at the singularity. They cannot linger near the horizon to alter $|A\rangle$ or $|B\rangle$ in the far future.
%\end{itemize}
%We will instead demonstrate --
%\begin{itemize}
%\item Back-reaction is bounded from below: \\
%Back-reaction cannot be tuned arbitrarily small without sacrificing the ability to distill $R_B$ from $R$.
%\item Back-reaction modulates the black hole evaporation: \\
%The back-reaction is not like sending signals into the black hole. Instead it modulates the evaporation process for a long time $\sim M^3$. Naturally it can alter the state of an old black hole including $|B\rangle$.
%\end{itemize}

%Note that for a practical paradox, it is essential to distill $|R_B\rangle$, or at least to have a system smaller than $R$ that entangles with $|R_B\rangle$ (such as performing a classical measurement). That is because Alice cannot bring all $\sim M^2$ qubits in $R$ back into the black hole without dramatically altering the geometry. She must process the information in $|R\rangle$ one way or another to reduce the amount of information she needs to carry. Our analysis applies to the distillation process and other measurement processes equally well.

The nature of the distillation process of course depends on how the information is encoded
in the Hawking radiation, which is determined by both the $S$-matrix and the initial state. For example if only the very first early quantum is entangled with the late quantum $B$, Alice can simply ignore all other early quanta. Such a trivial distillation needs not induce a meaningful back-reaction, but 
there is no reason to expect such a trivial entanglement structure would emerge out
of the $S$-matrix
\footnote{States with such a trivial entanglement structure
might form a complete basis in the Hilbert space of the Hawking radiation \cite{AveCho12,Cho13,Bou13}, 
but there is no reason to expect
the property of basis states to be carried over to a general
superposition. 
The fact that the Hilbert space for Hawking radiation is larger than the Hilbert space of the black hole \cite{Pag04}
suggests one should expect a general superposition rather than
a very special trivial state for the Hawking radiation.
}. In other words, our resolution does impose certain constraints on the black hole $S$-matrix. 
Spelling them out in more detail might lead to a better understanding of the $S$-matrix.
%Although they do not contradict the general expectations, they are still nontrivial and may lead to a better understanding of the $S$-matrix.

Let us briefly comment on the relation between our resolution to some existing proposals. The ER=EPR proposal of \cite{MalSus13} can be viewed as a physical way to enforce ``in=out'' by using the worm-hole geometry. This is compatible with part of our viewpoint: ``in=out'', i.e. complementarity, when there is no back-reaction. On the other hand,
when back-reaction does occur, our view is that it can occur by 
propagating signals through the normal part of the space-time, as opposed to through the Einstein-Rosen bridge.
The practical paradox may also be resolved by a limitation on the computation time of the distillation process \cite{HH}. Here we consider the possibility that this limitation can be somehow circumvented,
in which case an alternative resolution of the paradox is required.
Let us close this introduction by pointing out that 
our resolution is consistent with the possibility that Alice, with the powerful knowledge of the $S$-matrix and initial state, might be able manipulate the black hole to make some energetic late quantum i.e. what is commonly
referred to as the firewall. However a firewall has no reason to spontaneously develop during an unaltered evaporation process (one that suffers no back-reaction), so there is no violation of the equivalence principle.

{\it Outline}

In Sec.\ref{sec-back} we construct a model for the distillation process and the back-reaction.
In Sec.\ref{sec-AND} we show how ``complementarity + back-reaction'' should be enough to avoid the practical paradox. In addition, the firewall paradox has been formulated in various ways, and some of them do not involve an explicit distillation or measurement process. In Sec.\ref{sec-unit} we explicitly show that the resolution in one such formulation directly follows from our resolution to the practical paradox. This is not surprising since they are ultimately the same paradox. In Sec.\ref{sec-dis} we summarize our result and point out the possibility to decode the black hole $S$-matrix by performing computations on the nice slice.
%In addition to the practical paradox, the same information problem has been formulated as various requirements of new physics, such as firewalls \cite{AMPS}, hidden wormholes \cite{MalSus13}, frozen horizons \cite{Bou13a}, nonlocal dynamics \cite{Gid12}, final state quantum mechanics \cite{LloPre13}, and non-unitary evolutions within a causal patch. 

\section{Back-reaction}
\label{sec-back}

\subsection{The distillation process}

In order to address back-reaction, we need a more concrete description of how to extract information from the early Hawking radiation. Our argument focuses on a unitary distillation process, during which $R_B$ is separated from the rest of $R$. However it should be obvious that it applies to classical measurements, too. 
\footnote{By a classical measurement, we mean a non-unitary projection.} 
We will first argue for the following universal requirement for any distiller.

{\it A distiller that can extract $R_B$ from $R$ must carry some current that interacts with the Hawking radiation. Such a current is designed to match the expected pattern that encodes $R_B$ within $R$, which is determined by the black hole initial state, the $S$-matrix, and the specific late quantum $B$.}

In order to see this, we model the distillation process as the following unitary evolution: 
\begin{equation}
e^{-i\int H dtd^3x}|R\rangle|{\rm distiller\rangle}|0\rangle
=|R',{\rm distiller'}\rangle|R_B\rangle~,
\label{eq-scatter}
\end{equation}
where a time ordering is implied in the evolution operator. Initially, Alice has a memory stick of one empty qubit, where empty just means it has one unit of fixed and irrelevant information content. She plugs it into the distiller and together form the ``distiller system'', which interacts with the incoming Hawking radiation $R$. In the end this memory stick should be loaded with $R_B$ such that Alice can unplug it and bring it into the black hole. After this process $R$ loses the information in $R_B$ and becomes $R'$, which generically will be entangled with the distiller state that is also altered\footnote{For classical (projection) measurements, we can use exactly the same system but instead of the full quantum information of $R_B$, Alice is only allowed to carry away some classical information related to it.
}.

Eq.~(\ref{eq-scatter}) is in the Schr\"{o}dinger picture where the states evolve with time according to the full Hamiltonian density $H$. However only the coupling term $H_{\rm int}$ can be responsible for the transferring of information regarding $R_B$\footnote{We focus on the physical process represented by this Hamiltonian density, and assume that the distillation in Eq.~(\ref{eq-scatter}) is in-principle possible within a reasonable amount of time. As argued in \cite{HH}, the paradox can be resolved if the required time is too long, but we are looking for another resolution independent from that possibility.}:
\begin{equation}
H_{\rm int} = A_\mu J^\mu + \sum ({\rm radiation})({\rm current})~.
\end{equation}
The ``radiation'' operator acts on the state of Hawking radiation $R$ while the ``current'' operator acts on the state of the distiller system. Therefore $H_{\rm int}$ entangles them and transfers information. For simplicity let us focus on the standard interaction term in electromagnetism $(A_\mu J^\mu)$, while in general we expect
radiation of all types (e.g. gravitons, neutrinos, scalars) each of which is coupled to its corresponding current.
Our argument works in the same way, regardless of
the spin of the radiation particle.

Note that given the same initial pure state of the black hole, if we are interested in a different late quantum $B'$, we need to distill a different $R_{B'}$ correspondingly. Since the incoming radiation $R$ is still the same state, of course we need a different (initial) distiller state such that $J_\mu$ acts on it differently. 
Similarly, the requisite distiller state also depends on the initial state of the black hole.
If we have a different initial black hole pure state with the same macroscopic parameters, $R$ will be in a different state which encodes the information of $R_B$ differently. Aiming for the same $B$ according to a roughly identical evaporation process and classical geometry (for example the 24601st quantum), we will also need a different distiller state. 
In other words, the required initial distiller state depends on a number of things: it depends
on the classical label $B$ and on the state of the expected Hawking quanta, which in turn depends on the initial state of the black hole and the S-matrix. 

We find it convenient to express the dependence of both the
distiller state and the state of the incoming Hawking quanta
on these various features of the problem (the initial state of the black hole,
S-matrix, etc) using a somewhat unusual form of the Heisenberg picture.
In the usual Heisenberg picture, time evolution is transferred to the operators; 
states do not evolve. Here, we wish to go one step further: we encode
features of the initial states in the operators as well. 
It works as follows. Let $t=0$ be the time where the Schrodinger
picture and the (usual) Heisenberg picture agrees. In other words, we
say
\begin{eqnarray}
|\Psi_1 (t=0) \rangle^S = |\Psi_1 \rangle^{H'} \quad ; \quad
|\Psi_2 (t=0) \rangle^S = |\Psi_2 \rangle^{H'}
\end{eqnarray}
where the the superscripts $S$ and $H'$ denote the Schrodinger picture and
the (usual) Heisenberg picture respectively.
Here $|\Psi_1\rangle$ and $|\Psi_2\rangle$ denote two different
initial states. Let us further define a {\it modified} Heisenberg picture, denoted
by superscript $H$, as follows:
\begin{eqnarray}
|\Psi_1 (t=0) \rangle^S = |\Psi_1 \rangle^{H'} = U_1 |\Psi_0\rangle^H \quad ; \quad
|\Psi_2 (t=0) \rangle^S = |\Psi_2 \rangle^{H'} = U_2 |\Psi_0\rangle^H
\end{eqnarray}
where $|\Psi_0 \rangle^H$ is some common 'ancestor' state
(whose particular choice is not important) which are related to
$|\Psi_1 \rangle$ and $|\Psi_2 \rangle$ by the unitary transformations
$U_1$ and $U_2$ respectively. 
In our adaptation of the Heisenberg picture, the state is thus
always $|\psi_0\rangle^H$, independent of time and independent of
initial conditions. All the interesting information about the dynamics
and initial conditions are encoded in the operators:
\begin{eqnarray}
(A_\mu)_1^H &=& U_1^\dagger U^\dagger (t,0) (A_\mu)^S U (t,0) U_1~, \ \ 
(A_\mu)_2^H = U_2^\dagger U^\dagger (t,0) (A_\mu)^S U (t,0) U_2~, \\ 
(J_\mu)_1^H &=& U_1^\dagger U^\dagger (t,0) (J_\mu)^S U (t,0) U_1~, \ \ 
(J_\mu)_2^H = U_2^\dagger U^\dagger (t,0) (J_\mu)^S U (t,0) U_2~.
\end{eqnarray}
Henceforth, we will drop the subscript $1$ or $2$ which only
serves to remind us the operator in question, $A_\mu {}^H$ or
$J_\mu {}^H$, cares about the initial state. We will even drop the
superscript $H$ -- hereafter when we refer to operators, we mean
Heisenberg operators defined in this way. Therefore,
we write:
%More intuitively we can go to the Heisenberg picture and qualitatively denote this situation as
\begin{eqnarray}
H_{\rm int} &=& A_\mu({\rm black \ hole})J^\mu({\rm distiller})~, \\
A_\mu({\rm black \ hole}) &=& 
\left\{ S{\rm-matrix},~{\rm initial \ state}\right\}~, \\
J_\mu({\rm distiller}) &=& 
\left\{ A_\mu({\rm black \ hole},r_{\rm distiller}),S{\rm-matrix},B\right\}~.
\label{eq-detectorJ}
\end{eqnarray}
The content of the Hawking
radiation $A_\mu ({\rm black hole})$ depends on the black hole $S$-matrix and initial state. The current $J_\mu ({\rm distiller})$ required to distill $R_B$ depends on $B$ and also 
on the $S$-matrix which determines
how $R_B$ is encoded in $R$. 
There is a trivial dependence on $r_{\rm distiller}$ (location of the distiller) which determines at what time the radiation arrives at the distiller. 
%Finally, since $J_\mu$ is supposed to match the pattern in $A_\mu$ that hides the information of $R_B$, it should only be determined up to a pattern. Its amplitude should be somewhat adjustable as we denote by the constant $\alpha$.

%Note that this is a slight generalization of the usual Heisenberg picture. In the Schr\"{o}dinger picture if we are comparing between two different scenarios such as two black holes or distillers with different initial states, $|\psi_1\rangle^S$ and $|\psi_2\rangle^S$, we pretend that they were pre-evolved from one common ancestor state $|\psi_0\rangle$ by two different unitary operators.
%\begin{eqnarray}
%|\psi_1(t=0)\rangle^S = U_1|\psi_0\rangle^H~, \ \ 
%|\psi_2(t=0)\rangle^S = U_2|\psi_0\rangle^H~.
%\end{eqnarray}
%In our adaptation of the Heisenberg picture, the state is always $|\psi_0\rangle^H$, independent of time and across different scenarios. The pre-evolution is included in our definition of the Heisenberg operators:
%\begin{eqnarray}
%(A_\mu)_1^H &=& U^\dagger (t,0) U_1^\dagger (A_\mu)^S U_1 U (t,0)~, \ \ 
%(A_\mu)_2^H = U^\dagger (t,0) U_2^\dagger (A_\mu)^S U_2 U (t,0)~, \\ 
%(J_\mu)_1^H &=& U^\dagger (t,0) U_1^\dagger (J_\mu)^S U_1 U (t,0)~, \ \ 
%(J_\mu)_2^H = U^\dagger (t,0) U_2^\dagger (J_\mu)^S U_2 U (t,0)~.
%\end{eqnarray}
%This means that not only the time evolution, but also the possibly different initial states, are encoded in the operators. 
The advantage of our generalized Heisenberg picture is that it allows us to
use a language that is almost classical.
The operator $A_\mu ({\rm black hole})$ can be thought of as the
(state of) radiation from the black hole.
The operator $J_\mu ({\rm distiller})$ can be thought of as the (state of)
current of the distiller.
%is allowing us to treat the operators as the classical quantities. Although technically the Schr\"{o}dinger states $|R\rangle$ and $|{\rm distiller}\rangle$ are unlikely to be some eigenstates with definite classical $A_\mu$ and $J_\mu$, we can treat the Heisenberg operator $A_\mu({\rm black \ hole})$ literally as the (quantum state of) radiation from the black hole, and $J_\mu({\rm distiller})$ as the (quantum state of) current of the distiller. 
Occasionally, we would freely switch picture in our descriptions.
For instance, it should be obvious that ``changing the current $J_\mu({\rm distiller})$'' is in the 
generalized Heisenberg picture and ``changing the state $|{\rm distiller}\rangle$'' is in the Schr\"{o}dinger picture -- they could
even be the same change, expressed in different ways.

Given the current $J_\mu({\rm distiller})$, the discussion of back-reaction is straightforward. 
A current that can respond to incoming radiation is also a source itself. As we have focused on the EM part of the Hawking radiation, we can describe that (in Lorenz gauge) simply by
\begin{equation}
\partial_\mu\partial^\mu A_\nu({\rm distiller})=J_\nu({\rm distiller})~.
\end{equation}
Similar equations hold for other types of radiation, such as a massless scalar.
\footnote{
This type of equation assumes the particle corresponding to the radiation, absent interaction
with a current, is free. This does not hold, for instance, for gluons.
The story for such particles would be more involved, but
should
be similar in spirit to the one we are telling.}
The back-reaction we are interested in refers to how the black hole is affected by this $A_\mu({\rm distiller})$, or in general the field sourced by the distiller. We shall analyze it from two different perspectives. In Sec.\ref{sec-brane} we follow the membrane paradigm \cite{membrane} and treat the black hole as some object of size $\sim M$. In Sec.\ref{sec-nice} we analyze the Hawking process on the nice-slice \cite{Mat09}. Both pictures show that the back-reaction can modify the late Hawking quantum $B$.

%{\bf TO DO: Ultimately though, $H_{\rm int.} = 
%(A_\mu ({\rm black hole}) + A_\mu ({\rm distiller})) (J^\mu ({\rm black hole}) + J^\mu ({\rm distiller}))$.
%We should discuss this, especially noting that in our generalized Heisenberg picture,
%everything about the initial states of the black hole and distiller
%are in fact encoded in both $J^\mu ({\rm black hole})$ and $J^\mu ({\rm distiller})$
%(and likewise, $A_\mu ({\rm black hole})$ and $A_\mu ({\rm distiller}))$.
%The only difference between 
%$J^\mu ({\rm black hole})$ and $J^\mu ({\rm distiller})$ is location: 
%i.e. they simply denote the full $J^\mu$ localized respectively at the black hole
%and the distiller. And then, $A_\mu ({\rm black hole})$ and $A_\mu ({\rm distiller})$
%merely describes the $A_\mu$ sourced by each. Have to think about where to
%put these statements.
%}

\subsection{The membrane paradigm}
\label{sec-brane}

From the outside point of view, the black hole can be thought of as a membrane that sources Hawking radiation. Its effective current $J_\nu ({\rm black hole})$ can be worked out from:
\begin{equation}
\partial_\mu\partial^\mu A_\nu({\rm black \ hole})
=J_\nu({\rm black \ hole})~.
\label{eq-Jbh}
\end{equation}
This is a rather indirect way to describe the evaporation process.
%so in this section we cannot provide a concrete dynamics of this back-reaction. 
The analysis in the next section \ref{sec-nice} will give a more concrete picture. 
Here the goal is to put the distiller and the black hole on the same footing, 
and make plausible the notion that a non-trivial transformation of the distiller
(i.e. the distillation process) entails
a non-trivial transformation of the black hole (i.e. back-reaction).
%which makes it easier to show that the back-reaction cannot be tuned small and is in the right direction.

For the same reason that the distiller is influenced by the interaction between its own current and the radiation from the black hole (previously $H_{\rm int}$),
\begin{equation}
H_{\rm distiller} = A_\mu ({\rm black \ hole}) 
J^\mu({\rm distiller})~,
\label{eq-Hintdis}
\end{equation}
the black hole is also influenced by the radiation from the distiller,
\begin{equation}
H_{\rm black \ hole} = A_\mu ({\rm distiller}) J^\mu({\rm black \ hole}) \, .
\label{eq-Hintbh}
\end{equation}
This is the natural back-reaction of any distiller Alice wishes to employ. 
 
Our notation requires some explanation. From the quantum field theory
point of view, there is only one current operator
$J_\mu$ and one photon operator $A_\mu$. 
We can split $J_\mu$ into two halves, one non-zero only in
the space-time region containing the distiller i.e. $J_\mu ({\rm distiller})$, and
the other non-zero only in the region containing the black hole i.e. $J_\mu({\rm black \ hole})$.
The full $A_\mu$ sourced by the full $J_\mu$ can thus also
be split into two halves: 
$A_\mu({\rm distiller})$ sourced by $J_\mu ({\rm distiller})$ and 
$A_\mu({\rm black \, holes})$ sourced by $J_\mu ({\rm black \ hole})$. 
Out of the full product $A_\mu J^\mu$, 
Eqs. (\ref{eq-Hintdis}) and (\ref{eq-Hintbh}) are the cross-terms
which describe the influence of one system on the other.
The diagonal terms describe how each system evolves on its own,
which is not what we are interested in.

%Note that now we are not only in the Heisenberg picture, but also using the following natural convention. $J_\mu({\rm distiller})$ and $J_\mu({\rm black \ hole})$ are the current operator restricted to the regions occupied by the distiller and the black hole (the membrane) respectively. On the other hand the radiation operator $A_\mu$ usually hits the spacetime regions where the radiation sourced by both are present. Nevertheless in empty space $A_\mu$ from different sources combine linearly, and we have no trouble tracking them back to the sources. So $A_\mu({\rm black \ hole})$ and $A^\mu({\rm distiller})$ are still well-defined. Also note that the effect of self-coupling, such as $A_\mu({\rm black \ hole})J^\mu({\rm black \ hole})$, describes how the black hole would have been radiating without this back-reaction. That is given by our knowledge of the $S$-matrix and removed from this analysis. We focus on the additional change the black hole goes through in the presence of back-reaction.

It should already be obvious that Eq.~(\ref{eq-Hintdis}) and (\ref{eq-Hintbh}) are closely related. We can see that relation more clearly by integrating out $A_\mu$,
%up their contributions to the time evolution operator and 
utilizing the Green's function, $\mathcal{G^{\mu\nu}}$:
\begin{eqnarray}
\int dtdx^3~H_{\rm distiller} = 
\int dtdx^3 \int dt'dx'^3 & &
\mathcal{G^{\mu\nu}}({\rm from \ black \ hole \ to \ distiller}) 
\nonumber \\ \label{eq-Hdet}
& & J_\mu({\rm distiller})J_\nu({\rm black \ hole})'~, \\
\int dtdx^3~H_{\rm black hole} = 
\int dtdx^3 \int dt'dx'^3 & &
\mathcal{G^{\mu\nu}}({\rm from \ distiller \ to \ black \ hole}) 
\nonumber \\ \label{eq-Hbh}
& & J_\nu({\rm distiller})'J_\mu({\rm black \ hole})~.
\end{eqnarray}
In the first equation the $(x,t)$ integral goes over the distiller while the $(x',t')$ integral goes over the black hole. In the second equation it is the other way round. Presenting them side-by-side makes it clear that just like the black hole influences the distiller (first equation),
the distiller can influence the black hole (second equation).
The two expressions are not identical, however, due to the fact that
the retarded Green's function is not symmetric, 
because of causality.
%That is related to causality in the sense that black hole only receives %back-reaction at a later time than when it emits %radiation
%\footnote{Another potential difference comes from the time-ordering needed on $t$ but not on $t'$, which is more subtle but not directly relevant to our analysis.}.

The entire early Hawking radiation lasts a long time $\sim M^3$, which is usually much larger than the distance between the distiller and the black hole if Alice wishes to jump in later. Therefore the difference caused by the Green's function is small, and
Eqs.~(\ref{eq-Hbh}) and (\ref{eq-Hdet}) are expected to give fairly similar results.
We put forward the following conjecture:
%on the property of the black hole $S$-matrix:

{\it Given a typical black hole initial state and a typical late quantum $B$, the effective current $J_\mu({\rm black \ hole})$ that sources the Hawking radiation and the required current $J_\mu({\rm distiller})$ to distill $R_B$ -- both
depending on the S-matrix -- take forms which make
Eqs.~(\ref{eq-Hdet}) and (\ref{eq-Hbh}) comparable.}

The main substance of this conjecture is that 
nearly {\it all} the early radiation has to be processed by the distiller
(in order to obtain $R_B$), so the black hole is back-reacted for roughly the same duration as the distiller functions. This places some non-trivial constraint on the black hole S-matrix.
This conjecture implies that back-reaction cannot be tuned arbitrarily small. Recall that the state of Alice's memory stick has to be changed from the empty initial state $|0\rangle$ to the intended information content $R_B$. This change requires a transfer of information effected by $H_{\rm distiller}$, and a non-trivial change requires:
\begin{equation}
\left\langle \int dtdx^3~H_{\rm distiller}\right\rangle \gtrsim 1~.
\label{eq-bound}
\end{equation}
The conjecture then implies:
\begin{equation}
\left\langle \int dtdx^3~H_{\rm black \ hole}\right\rangle \gtrsim 1 \, ,
\end{equation}
meaning 
%Recall that the distiller current does have a tunable amplitude $\alpha$, and reducing it does reduce the back reaction. 
%\begin{equation}
%\alpha\propto\left\langle \int dtdx^3~H_{\rm black \ hole}\right\rangle~.
%\end{equation}
%However this also reduces the effect on the distiller which cannot be arbitrarily small. At the very least, the state of Alice's memory stick has to be changed from $|0\rangle$ to $|R_B\rangle$. This change requires a transfer of information therefore can only be achieved by $H_{\rm distiller}$, and the change is only real if
%\begin{equation}
%\alpha\propto
%\left\langle \int dtdx^3~H_{\rm distiller}\right\rangle \gtrsim 1~.
%\label{eq-bound}
%\end{equation}
%Otherwise, the uncertainty principle would forbid anyone from seeing the information transfer. In light of the similarity between $H_{\rm distiller}$ and $H_{\rm black \ hole}$, this means 
that information-wise, the change of the black hole state is similarly bounded from below.
%\footnote{Note that we have reverted back to the Schr\"{o}dinger picture to talk about the change in states. However Eq.~(\ref{eq-bound}) is about an expectation value which is picture independent.}. 

It should be stressed that we do not have a definitive proof that 
back-reaction is non-negligible. Indeed, we can be accused of assuming
the answer we want by positing the conjecture. Nonetheless,
the form of the black-hole-distiller interaction expressed in
Eqs. (\ref{eq-Hdet}) and (\ref{eq-Hbh}) is rather suggestive.
It suggests that the back-reaction on the black hole cannot be avoided
however clever the design of the distiller.
%At this point, we can see that several common objections to back-reaction are invalid. Usually the back-reaction is pictured as some ``instrumental'' factor, something we can reduce by a better design and has nothing to do with the problem of the information at hand. We have just shown this to be false. 
The source of back-reaction is the distiller current given by Eq.~(\ref{eq-detectorJ}). It is dictated by our desire to distill $R_B$, given the $S$-matrix and the initial state. Trying to reduce its back-reaction also diminishes the ability to distill $R_B$, thus there is a lower bound. 

It is worth emphasizing that the back-reaction is a back-reaction on the state of the black hole.
In other words, the radiation sourced by the distiller is not by itself
interesting -- indeed, that radiation enters, as opposed to emerges out of,
the black-hole horizon, and so appears to go in the wrong direction compared to
the direction of the late outgoing Hawking quantum $B$ -- it is interesting
only because the black hole is radiating to begin with i.e. the back-reaction is effected
by the {\it product} of the distiller and black hole currents. 
The physical effect of the back-reaction is an alteration of the black hole state, or 
can be thought of as 
a modulation of the evaporating process
%Another concern is that back-reaction is going in the wrong direction. Anything we send into the black hole should have crossed the horizon and ended up in the singularity very soon. They should not linger around the horizon to affect a late time quantum. We can dispel that concern, too. If the black hole was only passively receiving the flow of $A_\mu({\rm distiller})$ from the distiller, the effect should have depended on the flux $[A_\mu({\rm distiller})]^2\propto [J_\mu({\rm distiller})]^2$. Instead we showed that the back-reaction effect depends on the effective current of the black hole and would have been zero if the black hole was not radiating. The more appropriate physical interpretation of this back-reaction is a modulation of the original evaporating process
\footnote{This is the general interaction between two sources emitting the same type of radiation \cite{us}. The leading order effect is not that of one sending something to another, but a mutual modulation: one source appears to be emitting more or less under the influence of the other. This is equivalent to an interference effect between their emissions. In certain cases the original emission greatly amplifies the influence of the incoming radiation and serves as a good detector \cite{HuiMcW12}.}. {\it If Alice performs distillation for a long duration $\sim M^3$, the black hole's emission is also modulated by a comparable duration, so the state of a particular late quantum is accordingly modified.} The same physical interpretation will emerge from our analysis in the next section.

\subsection{The nice-slice}
\label{sec-nice}

There is another way to understand how $A_\mu({\rm distiller})$ affects the evaporation process. Instead of treating the black hole as a source with an effective current, let us follow another description of the Hawking radiation. We construct the ``nice slice'' which are space-like surfaces going through the horizon \cite{Mat09}. Curvature remains small everywhere, while the near horizon region has an expanding geometry. We can use quantum field theory to describe the comoving modes whose wavelengths are being stretched by the expansion. When the physical wavelength of a comoving mode is short, it should be in the vacuum state of a locally flat region. Without outside influence, it stays in that comoving vacuum state while being stretched. As the wavelength becomes longer and the modes migrate into the asymptotic region of the nice slice, this comoving vacuum disagrees with the asymptotic vacuum. This pair-production in the expanding background is the source of Hawking radiation in the outgoing modes.

\begin{figure}[ht]
\begin{center}
\includegraphics[width=8cm]{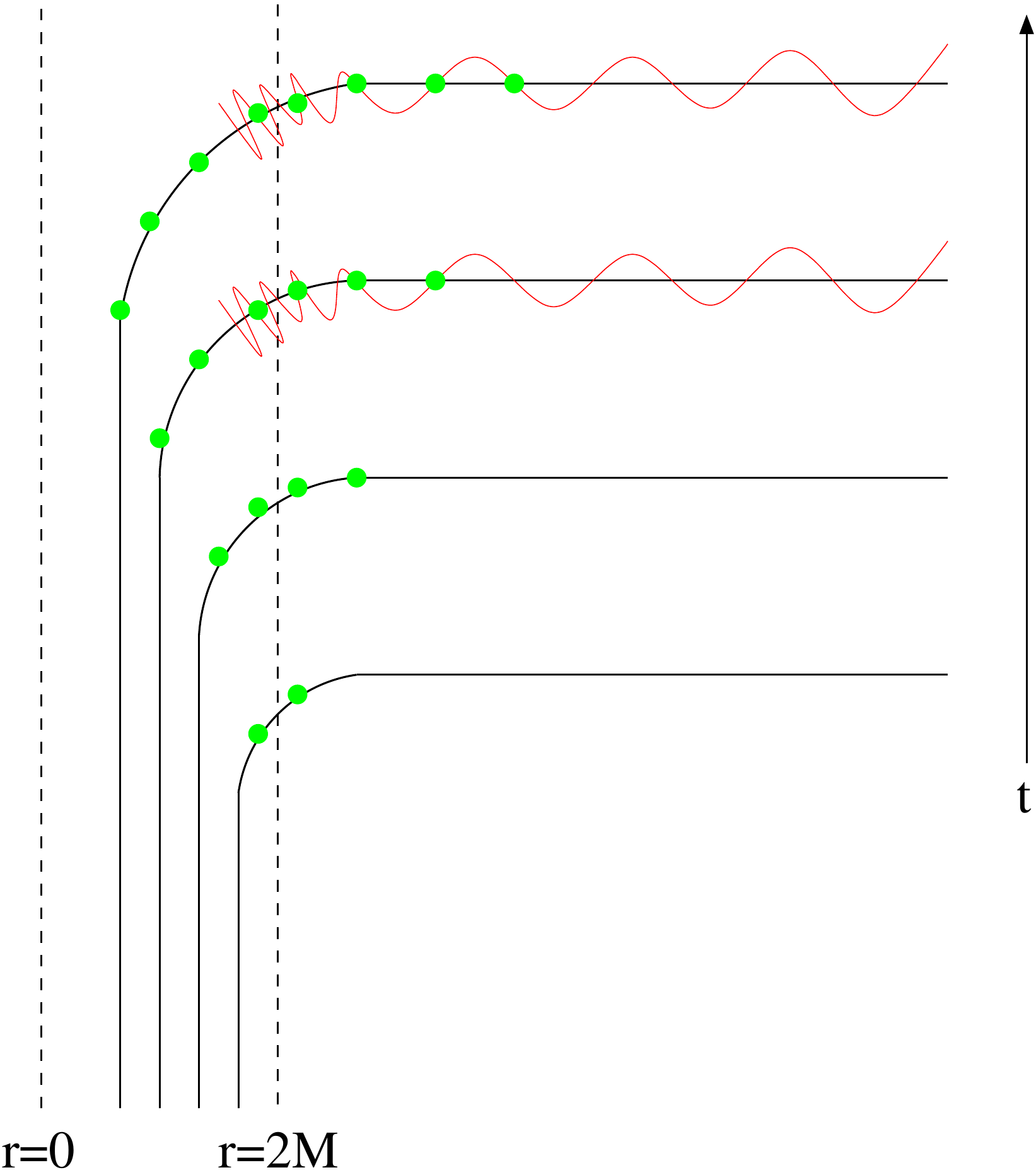}
\caption{The green dots represent pair-produced particles in the near horizon region which are being pulled apart by the expanding geometry. The
horizontal and vertical directions represent schematically the
Schwarzschild $r$ and $t$. The first two (lower) slices show the pair-production from vacuum, which leads to the Hawking radiation from an unaltered evaporation process. The second two (higher) slices are the same process under the effect of back-reaction from a distiller, which fills the near horizon region with $A_\mu({\rm distiller})$ drawn as the red waves. They are blue-shifted and piled up near the horizon. The pair-production is no longer from vacuum, so the evaporation process is altered. 
\label{fig-nice}}
\end{center}
\end{figure}

When Alice operates her distiller, the above stretching process proceeds with the near horizon region permeated by $A_\mu({\rm distiller})$. As depicted in Fig.\ref{fig-nice}, this modifies the dynamics. The comoving modes are affected by $A_\mu({\rm distiller})$ during the stretching process and do not have to stay in the comoving vacuum. 
%Alternatively we can understand this effect in the asymptotic region. If we expect the asymptotic regions to start in vacuum, then the near-horizon stretching process disagrees with it and of course creates particles. However now we expect the asymptotic region to be filled with $A_\mu({\rm distiller})$. A generic ``disagreement'' with such non-vacuum state sometimes creates particles, and sometimes annihilates particles. 

%This echoes the description in the previous section that the back-reaction on the black hole is a modulation effect. It is sometimes constructive and sometimes destructive, therefore a general modification to the evaporation process. 
Note that on the nice slice, the objection of a ``wrong direction'' (see discussion in
Sec. \ref{intro}  and \ref{sec-brane}) does not hold. The proper time progresses very slowly in the interior portion of the nice slices. So $A_\mu({\rm distiller})$ actually cannot cross the expanding near-horizon region and indeed linger around to modify the later Hawking radiation. This leads to a spectrum different from the original Hawking radiation. {\it The current $J_\mu({\rm distiller})$ to distill $R_B$ leads to an external influence $A_\mu({\rm distiller})$ that constantly modulates the evaporation process by acting on the comoving modes near the horizon in the nice-slice.} The process of how $A_\mu ({\rm distiller})$ effectively alters the state of the late radiation should be treatable with standard field theory techniques. The problem is fairly analogous to computing inflationary perturbations under the influence of disturbances that cause excitations above the Bunch-Davies state.

\section{Complementarity Prevails}
\label{sec-AND}

In this section we go over a few different scenarios and demonstrate how the effect of back-reaction can 
help resolve the practical paradox. They suggest that complementarity is self-consistent,
if back-reaction is correctly taken into account. 

\subsection{Bob's picture}
\label{sec-Bob}

First let us consider Bob, who has no intention, nor possibility, of jumping into the black hole. He can check whether the evaporation process is unitary while staying very far away, $r_{\rm Bob}\gtrsim M^3$. He will operate his distiller with the current given by Eq.~(\ref{eq-detectorJ}). The black hole will not be there when the back-reaction from Bob arrives, so he has no reason to worry about anything we said in the previous section. When the radiation and back-reactions cross path in empty regions away from both the black hole and the distiller, there is basically no interaction. After distilling $R_B$ from $R$, Bob will receive the late quantum $B$ and confirms a unitary evaporation process. 

\subsection{A careless Alice}
\label{sec-Alice1}

Alice would like to replicate Bob's process at a closer distance $r_{\rm Alice}\ll M^3$, such that she can carry the result $R_B$ into the black hole. She might believe that a simple modification 
of Bob's distillation is enough. Before the early quanta $R$ reach Bob, she can intercept them with her own distiller with the current also given by Eq.~(\ref{eq-detectorJ}). The difference in the variable $r$ takes care of the unitary transformation between the state of $R$ at her position and Bob's, so she expect to get the same result $R_B$.

Following this simple logic, Alice's plan is doomed to fail. Since $r_{\rm Alice}\ll M^3$, the back-reaction from her distiller starts to affect the black hole soon after distillation begins, and continues through almost the entire process. Since the state of the black hole is altered, it will not emit the same Hawking radiation. The difference might be small, but the distillation is a very delicate process where $J_\mu({\rm distiller})$ is designed to match a particular content of $A_\mu(\rm black hole)$. Using the same current on a different radiation content, both $B$ and $R_B$ will be modified to something totally unexpected, and her check of unitarity will simply fail.

\subsection{A careful Alice}
\label{sec-Alice2}

Alice can try to be more careful. Now that she knows about the back-reaction, she realizes that distilling $R_B$ given by an unaltered evaporation process is a fool's errand at her location. Her action changes the black hole and she needs to take that into account. In principle, she can do a much more involved calculation to keep track of how both the black hole and her memory stick are simultaneously affected. She can then construct a device that manipulates both systems for a long time $\sim M^3$. The information in a late quantum $B$ will not be the same as the naturally evaporating black hole, and it can be entangled with her memory stick.

Assuming that Alice performs the calculation correctly, she can indeed confirm the desired entanglement. However, this entanglement has little to do with the $S$-matrix of the unaltered evaporation of this black hole. Alice is simply checking the result of her manipulation. It may be surprising, but not against any physical law that the outcome is dynamically enforced. Since she deliberately manipulated the state of $B$, there is no reason that it is still purified by $A$\footnote{In principle the state of $A$ is also changed by what Alice did, but the exact change does not matter in our argument. Alice enforced an observable entanglement on $B$ and $R_B$, which must exclude another observable entanglement.}. So when she jumps in, she will see those high-energy quantum she created. This comes from her impressive ability to guide the black hole through a very special evaporation process during a long time $\sim M^3$, and does not violate the equivalence principle or any other low energy physical laws.

\subsection{Combining Alice and Bob}
\label{sec-combine}

Now let us have both Alice and Bob. Alice still follows her fixed trajectory to cross the horizon at some time, say $t_{\rm cross}$. However she will not attempt to do anything to the Hawking radiation. If she did, the black hole would be altered by back-reaction, and neither she nor Bob would have access to an unaltered evaporation process. Alice will just fall through the horizon and check equivalence principle, namely the smoothness of horizon as the entanglement between the interior mode $A$ and the exterior mode $B$.

Bob is in charge of checking the unitary evaporation. He knows about Alice's schedule, and agrees to check the state of a near-horizon mode $B$ at $t_{\rm cross}$. He operates his detector to distill $R_B$, and then later observes $B$ directly as a quantum in the late Hawking radiation. Now let us examine if the experiences of Alice and Bob can contradict each other.

Let Bob starts at $r_{\rm Bob}(t = t_{\rm cross}/2)=t_{\rm cross}/2$, such that the back-reaction cannot reach the horizon before Alice falls through. This way Bob's distiller cannot causally change Alice's experience. If Bob just stays there, he can eventually see $B$ as a late quantum and confirm that an unaltered evaporation is unitary. Although he is slightly closer to the black hole than the ideal distance described in Sec. \ref{sec-Bob}, most of the back-reaction signals still cross path with the late quantum mostly in empty space. We should allow Bob to use the trivial $J_\mu$ in Eq.~(\ref{eq-detectorJ}) and treat $B$ as unaltered. So, staying at this large distance, Bob should confirm the entanglement between $B$ and $R_B$ given by the unaltered black hole $S$-matrix\footnote{Hawking quantum coming out at even later times are modified by the back-reaction.}. Now Bob's observation is in conflict with the normal horizon experienced by Alice. Fortunately they cannot compare their results (if Bob stays put), so by the assumption of complementarity this is not a paradox.

\subsubsection{Fast approaching Bob}
\label{sec-Bobfall}

\begin{figure}[ht]
\begin{center}
\includegraphics[width=8cm]{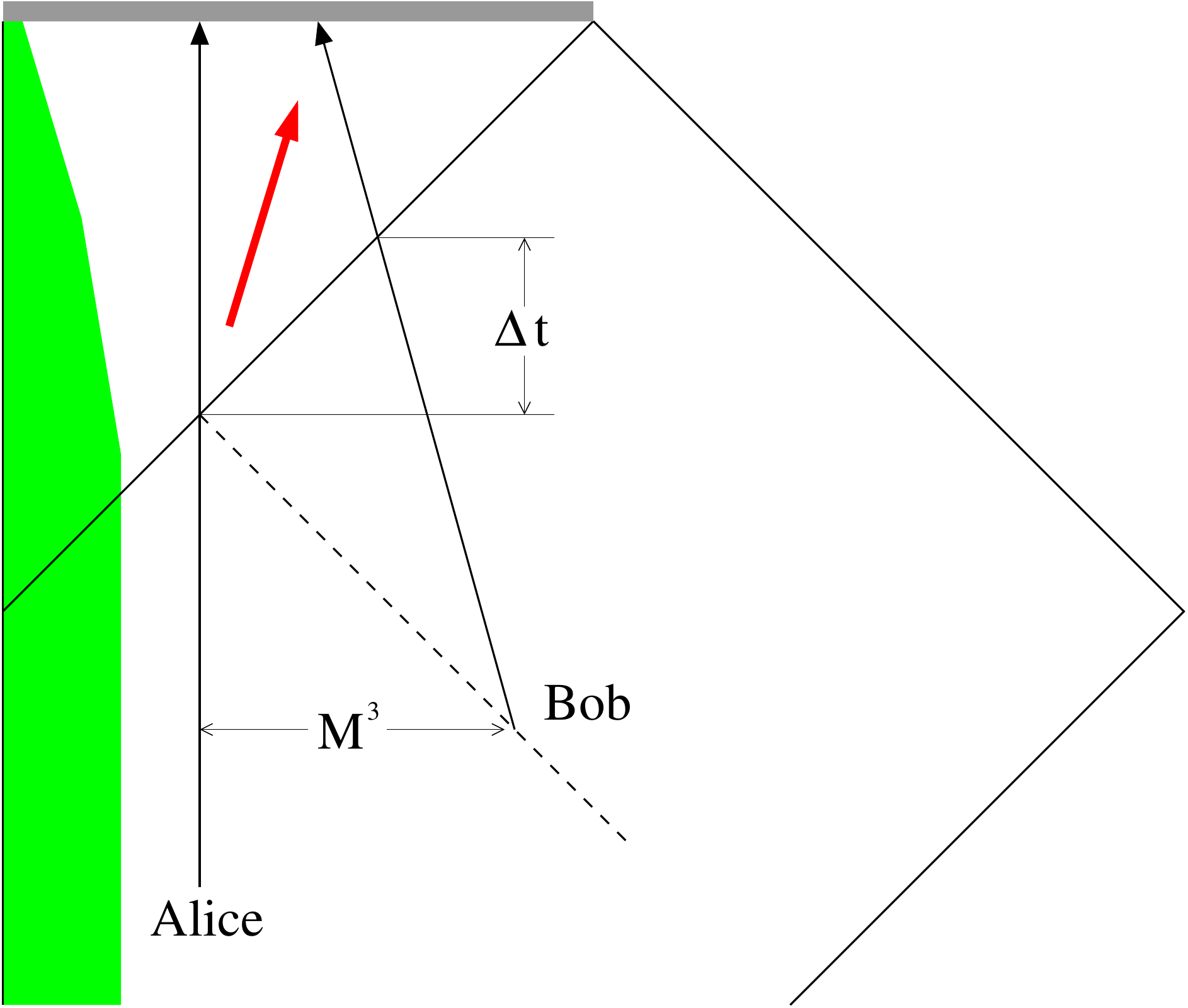}
\caption{Bob starts to distill $R_B$ from $R$ outside the past lightcone of the point which Alice crosses the horizon, which means that he has to be $\sim M^3$ away in the beginning. He approaches the black hole and follows Alice into the horizon $\Delta t$ later. If $\Delta t$ is too small, the early Hawking quanta have Planckian wavelength such that Bob cannot read them. If $\Delta t$ is too big, then the message from Alice (the red, thick arrow) will be Planckian such that Bob cannot read it.   
\label{fig-Bobfall}}
\end{center}
\end{figure}

In order for Alice and Bob to compare their results, Bob cannot stay this far all the time. He has to approach the black hole very fast and cross the horizon at some $t<t_{\rm cross}+\Delta t$. Note that the $t$ defined here are the Schwarzschild time when the observers cross some fixed distance outside but near the horizon, for example $r=2M + 1cm$. Depending on $\Delta t$, there is a chance that he can still communicate with Alice. This situation is drawn in Fig.\ref{fig-Bobfall}.

The first constraint on such a scenario is that the Hawking radiation should not be too blue-shifted for Bob.
\begin{eqnarray}
v &=& \frac{M^3}{M^3+\Delta t}\approx 1-\frac{\Delta t}{M^3}~, \\
\gamma &\sim& \sqrt{\frac{M^3}{\Delta t}}~, \\
E_{R, {\rm Bob}} &\sim& M^{-1}\gamma = \sqrt{\frac{M}{\Delta t}}~.
\label{eq-BobR}
\end{eqnarray}
Bob would like the Hawking quanta to remain sub-Planckian in his frame, therefore $\Delta t$ needs to be much larger than $M$. This, on the other hand, makes Alice's job quite difficult. First of all, there is a geometric constraint that if Bob jumps in $\Delta t$ later than Alice, then Alice has to send a message within 
\begin{equation}
\Delta \tau_{\rm Alice} < M e^{-\Delta t/M}~,
\end{equation}
such that Bob can receive it before crashing into the singularity \cite{SekSus08}. Note that $\Delta \tau$ here refers to the proper time for Alice, and the exponential relation between this time and the Schwarzschild time is the key of this argument. She will only have $\Delta \tau$ to explore the black hole interior before sending the information, so the wavelength of the photon she sends that can contain the information about the interior mode $A$ must be bounded by the same quantity.
\begin{equation}
\lambda_{\rm Alice}<\Delta \tau_{\rm Alice}~.
\end{equation}
This message will again be somewhat blue-shifted to Bob.
\begin{equation}
E_{\rm message \ from \ Alice} 
= \frac{\gamma}{\lambda_{\rm Alice}}
>\sqrt{\frac{M^3}{\Delta t}}M^{-1}e^{\Delta t/M}
=\sqrt{\frac{M}{\Delta t}}e^{\Delta t/M}~.
\label{eq-BobAlice}
\end{equation}
We have used the same $\gamma$ as in Eq.~(\ref{eq-BobR}), and here it is only an approximation. However, the point is that when $\Delta t>M$, the exponential behavior dominates and the exact $\gamma$ will not compensate its effect. We see that there is no way to make Eq.~(\ref{eq-BobR}) and (\ref{eq-BobAlice}) simultaneously small. Bob either has to deal with Planck scale Hawking radiation, or a Planck scale message from Alice. {\it If we want to avoid back-reaction, there is a fundamental obstruction to observing $A$ and $R_B$ together.}

\subsection{Unitary evolution}
\label{sec-unit}

The practical paradox (Alice in Fig. \ref{fig-intro}) is the strongest version of the AMPS argument. Following the footsteps of how we resolved it, the resolution to other versions becomes transparent. For example, one way to phrase the AMPS paradox requires no distillation since it only discusses the validity of a quantum mechanical description. Let $|\phi\rangle$ be the state of an infinite spatial slice before its matter collapses into a black hole. We can follow the quantum mechanical evolution to a later time where $A$, $B$ and $R$ are all present. In order to maintain a normal horizon between $A$ and $B$, and a unitary evaporation process relating $B$ and $R$, $A$ and $R$ must contain the same information. The information paradox can be stated as the violation of the unitarity evolution from $|\phi\rangle$ to $|A\rangle|B\rangle|R\rangle$. 

\begin{figure}[ht]
\begin{center}
\includegraphics[width=8cm]{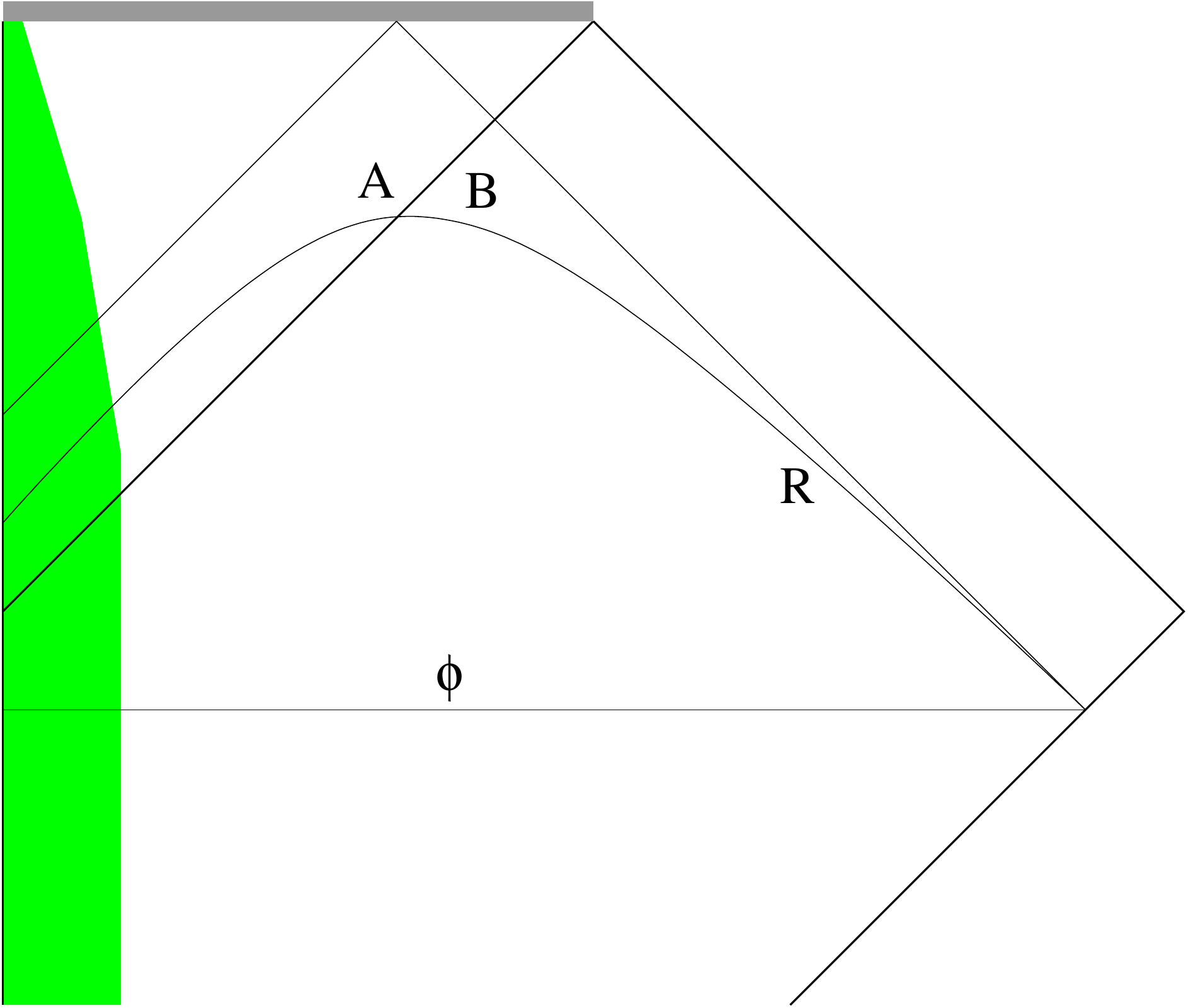}
\caption{The attempt to fit the interior partner $A$, the late quanta $B$, and early Hawking radiation $R$ into one causal patch. Inevitably either $A$ or $R$ has to be a Planckian quantity and stops us from using low energy quantum mechanics to evolve from $|\phi\rangle$ to $|A\rangle|B\rangle|R\rangle$.  
\label{fig-fit}}
\end{center}
\end{figure}

Complementarity claims that this is not a paradox, unless the spatial surface describing this evolution can fit into the causal patch of some observer. Usually we picture the situation that $R$ is very far away and cannot fit into the same causal patch with $A$ so it is not a problem. One might instead try to fit $A$, $B$ and $R$ into
one causal patch as is shown in Fig.\ref{fig-fit}. However following the discussion of a fast-approaching-Bob in Sec.\ref{sec-Bobfall}, we can see that such a fit is problematic. The top of this causal patch can be treated as where a fast-approaching Bob eventually ends up. Thus the early radiation $R$ on this slice is exactly the one Bob reads on his journey, and its wavelength is given by Eq.~(\ref{eq-BobR}). For the same reason, the physical size of $A$ on this slice is related to the message sent by Alice as in Eq.~(\ref{eq-BobAlice}). So following the math in the previous section, either the interior mode $A$ or the early radiation $R$ has to be a Planckian quantity. Complementarity is intrinsically a statement on low energy physics, so there is a natural and implicit fine-print: {\it Not only do we need $A$, $B$ and $R$ to fit into a causal patch, all of them should be low energy quantities.} This is violated here, thus cannot qualify as a paradox for complementarity. A more thorough analysis of this simple geometrical fact can be found in \cite{IlgYan13}.

\section{Discussion}
\label{sec-dis}

We argue that the practical paradox formulated by AMPS \cite{AMPS} can be resolved without new physics. A quick summary of this practical paradox is for Alice to 
\begin{enumerate}
\item distill $R_B$ from the early Hawking radiation $R$,
\item bring $R_B$ to check its entanglement with the corresponding late near horizon mode $B$,
\item cross the horizon to check the entanglement between $B$ and its interior partner $A$. 
\end{enumerate}
We show that the ``distillation'' process, if carried out in the causal past of the point Alice jumps in, 
modifies the black hole through back-reaction. We make a plausible argument that
a successful (non-trivial) distillation implies a minimum level of back-reaction.

In light of this, there are two possible outcomes for Alice and neither is paradoxical. If Alice is not careful enough and believes that she can distill $R_B$ from $R$ just like Bob (who is $\sim M^3$ away) does, her check of unitary evaporation will fail. That is because the back-reaction makes the black hole and her distiller interact. Without taking this into account, the subtle entanglement she tries to verify would not exist. She can perform the distillation also from $\sim M^3$ away to prevent back-reaction (i.e. the Bob situation). 
In that case she will confirm the unitary evaporation but will be unable to jump in and observe $A$, thus no practical paradox.

If Alice is very careful, she can stay close and perform something that is much more involved than a simple distilling. She can design a device to coherently affect the black hole together with a qubit she holds. This ensures that the particular late outgoing quantum does end up being maximally entangled with her qubit. However doing so is exactly using her powerful knowledge to disrupt the entanglement between $A$ and $B$. While crossing the horizon, Alice will see some high energy quanta due to the mismatch in $A-B$, but it is not a spontaneous violation of the equivalence principle. Alice has been manipulating the black hole for a long time and eventually has to taste her own medicine. In short, Alice herself creates the firewall.

This back-reaction is causal, and we present a model of how it works using the standard theory of radiation-source response. Furthermore, we argue that, viewed in the nice slices, breaking the cross-horizon entanglement is the natural interpretation of the physical effect of this back-reaction. 
%Thus our description does not require any new physics. It also does not modify the gedanken experiment if $R_B$ is distilled outside causal contact to the black hole. Since there was no paradox in that setting, any modification would have been redundant. 
Our resolution of the AMPS paradox requires no new physics, and is perhaps the most conservative one proposed so far. The starting point of our argument is a very general statement: Eq.~(\ref{eq-scatter}) must effectively describe any practical method to distill $R_B$ from $R$. For a more detailed implementation, we picked the least intrusive method, that the distiller simply waits for the Hawking radiation to come. Since there is already back-reaction in such a minimal setup, it seems unlikely that more intrusive/elaborated methods, such as mining \cite{Mining,Bro12}, can evade back-reaction.

In fact we would like to make a stronger statement by employing the following general definition of the distillation process. Whenever and wherever Hawking radiation stops free-streaming in vacuum and interacts with something else, those spacetime regions are considered part of the ``distiller''\footnote{For example one can try to arrange mirrors that reflect radiations to shield the black hole from the back-reaction. These mirrors will inevitably affect the normal out-going radiation as well
-- mirrors that perfectly transmits in one direction and reflects in the other don't exist.
We can just treat them effectively as part of the distiller.
There is a proposed set-up that includes an auxiliary system \cite{AMPSS} which has no clear interpretation purely in the bulk, so we do not consider it as constituting a practical paradox.}.
Eq. (\ref{eq-scatter}) continues to describe such a distillation process, and our back-reaction argument
should apply.

Although our resolution is conservative, it has interesting implications.
We provide two qualitative descriptions of how the back-reaction modulate the evaporation process in Sec.\ref{sec-back}, according to the membrane paradigm and the expanding geometry on the nice slices. The latter picture is particularly interesting, since in principle the calculation is similar to the one carried out in the context of inflation, in the presence of external disturbances that cause non-trivial excitations.
%the effective field theory of inflation where many techniques are known %\cite{AlbKal02,BurCli03,CirdeV04,CheCre07,WeiS08}.
It is possible that we can employ those techniques and try to learn something about the black hole $S$-matrix.

\acknowledgments

We thank Raphael Bousso, Ben Freivogel, Daniel Harlow and Robert Myers for useful discussions. The work of Lam Hui is supported by the DOE and NASA under cooperative agreements DE-FG02-92-ER40699 and NNX10AN14G. He thanks Gary Shiu and Henry Tye
at the IAS at the Hong Kong University of Science and Technology for hospitality. The work of I-Sheng Yang is supported in part
by the Foundation for Fundamental Research on Matter (FOM), which is part of the Netherlands Organization for Scientific Research (NWO).

\bibliographystyle{utcaps}
\bibliography{all}

\end{document}